\documentclass[showpacs,aps,prl,twocolumn,superscriptaddress]{revtex4-1}
\usepackage{graphicx} 
\usepackage{dcolumn}
\usepackage{bm}
\usepackage{amssymb,amsmath}
\usepackage{epstopdf}
\usepackage{color}
\def\etal{$\it{et~al.}$}

\begin{document}
\title{Magnetoplasmons in quasi-neutral epitaxial graphene nanoribbons}

\author{J. M. Poumirol}
\email{poumirol@magnet.fsu.edu}
\affiliation{National High Magnetic Field Laboratory, Tallahassee, FL 32310, USA}
\author{W. Yu}
\affiliation{School of Physics, Georgia Institute of Technology, Atlanta, GA 30332, USA}
\author{X. Chen}
\affiliation{School of Physics, Georgia Institute of Technology, Atlanta, GA 30332, USA}
\author{C. Berger}
\affiliation{School of Physics, Georgia Institute of Technology, Atlanta, GA 30332, USA}
\affiliation{CNRS/Institut N\'{e}el, BP166, 38042 Grenoble, France}
\author{W. A. de Heer}
\affiliation{School of Physics, Georgia Institute of Technology, Atlanta, GA 30332, USA}
\author{M. L. Smith}
\affiliation{Sandia National Laboratories, Albuquerque, NM 87185, USA}
\author{T. Ohta}
\affiliation{Sandia National Laboratories, Albuquerque, NM 87185, USA}
\author{W. Pan}
\affiliation{Sandia National Laboratories, Albuquerque, NM 87185, USA}
\author{M. O. Goerbig}
\affiliation{Laboratoire de Physique des Solides, CNRS UMR 8502, Univ. Paris-Sud, F-91405 Orsay cedex, France}
\author{D. Smirnov}
\affiliation{National High Magnetic Field Laboratory, Tallahassee, FL 32310, USA}
\author{Z. Jiang}
\email{zhigang.jiang@physics.gatech.edu}
\affiliation{School of Physics, Georgia Institute of Technology, Atlanta, GA 30332, USA}

\date{\today}

\begin{abstract}
We present infrared transmission spectroscopy study of the inter-Landau-level excitations in quasi-neutral epitaxial graphene nanoribbon arrays. We observed a substantial deviation in energy of the $L_{0(-1)}$$\to$$L_{1(0)}$ transition from the characteristic square root magnetic-field dependence of two-dimensional graphene. This deviation arises from the formation of upper-hybrid mode between the Landau level transition and the plasmon resonance. In the quantum regime the hybrid mode exhibits a distinct dispersion relation, markedly different from that expected for conventional two-dimensional systems and highly doped graphene.
\end{abstract}


\maketitle

Graphene plasmons, the collective oscillations of Dirac fermions, have recently attracted a great deal of attention \cite{Ju,Xia,Crassee,Koppens_N12,Basov_N12,Novoselov}. It has been theoretically suggested that graphene may be able to replace noble metals in future plasmonic devices, as it can host surface plasmons with higher degree of confinement and longer lifetime, owing to its extraordinary material properties \cite{Jablan,Koppens,Engheta}. Moreover, recent advances in graphene synthesis and fabrication have made it possible to pattern large-scale micrometer-sized structures with increasing carrier mobility and tunable carrier density \cite{deHeer,Pan,Park}. These advances enable broadband graphene plasmonics operating at terahertz (THz) frequencies and make graphene promising candidate material for next generation optoelectronics.

The ability to enhance light-matter interactions in graphene via confined plasmons has been demonstrated recently in graphene ribbons \cite{Ju,Yan1}, disks \cite{Xia,Yan2}, and tapered nanostructures \cite{Koppens_N12,Basov_N12}. Experimentally, surface plasmons manifest themselves as resonance absorptions in the transmission spectra, and the resonance frequency ($\omega_{pl}$) can be tuned by varying external parameters such as the graphene dimension and/or the carrier density ($n_{el}$). For graphene micro-ribbon arrays, for example, Ju \etal\ \cite{Ju} have shown that $\omega_{pl}$ depends on the width of the ribbon ($\omega_{pl}$$\propto$$W^{-1/2}$) and the Fermi energy ($\omega_{pl}$$\propto$$\sqrt{E_F}$$\propto$$\sqrt[1/4]{n_{el}}$), where the $\omega_{pl}$$\propto$$\sqrt[1/4]{n_{el}}$ dependence is unique to graphene. Previously reported results on graphene plasmons were mostly obtained in highly doped samples with $E_F$$\geq$300 meV \cite{Yan1,Xia,Koppens_N12,Yan2,Crassee}. Investigations near the charge neutrality point, where interaction-induced deviations from the Fermi-liquid picture are expected \cite{Neto12}, are still missing. In this particular regime, large wave vector (\textit{e.g.}, $q$$\sim$$\pi/100$ nm$^{-1}$) graphene plasmons are difficult to probe at zero magnetic field. High-mobility graphene specimens are needed to achieve exceptional optical field confinement and long plasmon lifetime.

In this Letter, we report on the observation of plasmon-type collective excitations in quasi-neutral epitaxial graphene nanoribbon (GNR) arrays exposed to a perpendicular magnetic field. Most saliently, the scaling behavior as a function of the wave vector ($q$) and the magnetic field ($B$) allows us to identify this mode with the upper-hybrid mode (UHM) between the plasmon resonance and the $L_{0(-1)}$$\to$$L_{1(0)}$ Landau level (LL) transition. This scaling is different from that of the UHM in conventional two-dimensional electron gases (2DEGs) with parabolic bands or in highly doped graphene as well as from that of magnetoexcitons. Furthermore, we show the possibility to confine plasmons in narrow GNRs allows to probe the dispersion relation of the UHM in a large parameter range. For the 100nm-wide GNR arrays, we observe a wavelength shrinkage of $\sim$165, a value difficult to achieve in common plasmonic materials, but in agreement with that predicted for graphene \cite{Jablan}.

The GNR array samples were fabricated from multilayer epitaxial graphene (MEG) grown on the C-face of SiC. The first few layers of graphene close to the SiC-graphene interface are highly doped due to charge transfer from SiC, while the subsequent top layers are practically charge-neutral and behave like isolated graphene monolayers \cite{Sadowski,Hass,Orlita,Sprinkle}. Three series of GNR arrays (200~nm, 100~nm, and 50~nm ribbons width) were patterned via electron-beam lithography in a large writing field, followed by oxygen plasma etching and finally vacuum annealing at 600 $^{\circ}$C for 2 hours to remove resist residues. The length of each ribbon is 400 $\mu$m. The inset in Fig. 1(a) shows an atomic force microscopy (AFM, Park Systems XE-100) image of a 100nm-wide GNR array sample. The GNR width/gap ratio is $W/d$=1:1, unless otherwise noted in the discussion below.

\begin{figure}[t]
\includegraphics[width=8cm]{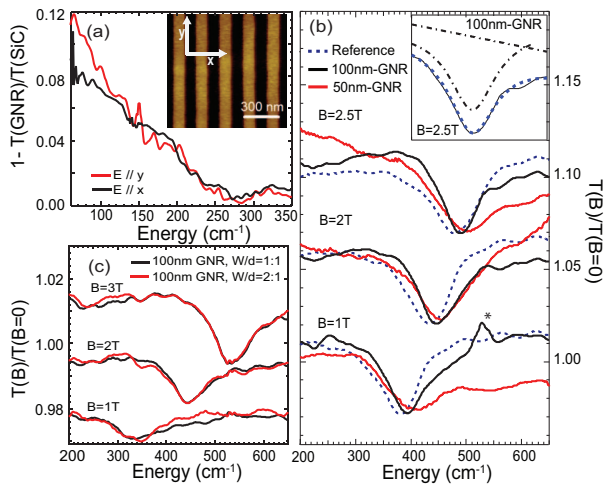}
\caption{(color online) (a) Polarized, zero-field relative transmission spectra, 1-$T_{\text{GNR}}/T_{\text{SiC}}$, of a 100nm-wide GNR array sample. Inset: AFM image of the sample. (b) Main panel: Normalized magneto-transmission spectra of 2D graphene (blue), 100nm-wide (black) and 50nm-wide (red) GNR arrays measured with un-polarized IR light and at different magnetic fields. The star symbol points to a $B$-independent spectral artifact associated with 60 Hz harmonics. Inset: Example of single Lorentzian fit (blue dash line) to the data (thin black line) with a slightly tilted baseline. (c) Comparison of two 100nm-wide GNR array samples with different width/gap ratios: $W/d$=1:1 (black) and $W/d$=2:1 (red). Spectra are scaled and offset vertically for clarity.}
\end{figure}

Infrared (IR) transmission spectroscopy measurements were performed at 4.2 K using a Bruker IFS 113v Fourier-transform IR spectrometer. The radiation from a mercury lamp was delivered to the sample via evacuated light-pipes, and the intensity of the transmitted light was detected by a composite Si bolometer. A $\sim$1~mm$^2$ square aperture was placed on the sample surface to reduce the amount of stray IR light around the GNR array.

Figure 1(a) shows the normalized zero-field transmission spectra of a 100nm-wide GNR array with the IR light polarization parallel or perpendicular to the ribbon direction. For both polarizations, no plasmon-related spectral features are observed. Instead, the spectra exhibit a Drude-like behavior with a zero-energy absorption peak, possibly due to unintentional growth of 2D graphene on the back side (Si-face) of the SiC substrate \cite{note_Raman}. The absence of plasmon resonance peaks at $B$=0 is likely due to the low carrier concentration in our GNRs, leading to very small oscillator strength for the plasmon resonance.

In order to probe plasmons in quasi-neutral GNR arrays, we apply a magnetic field perpendicular to the graphene plane and force electrons into cyclotron resonance (CR). The plasmon mode then couples with the CR, forming an UHM with energy \cite{Chiu,Demel,Goerbig_11}
\begin{equation}
\hbar\omega_{uh}(q)\simeq \sqrt{\Delta_{n_F}^2 + \hbar^2\omega_{pl}^2(q)},
\end{equation}
where $\hbar$ is Planck's constant, and $\Delta_{n_F}$ is the LL transition energy between the last occupied LL, $n_F$, and the first unoccupied one, $n_F$+1. For quasi-neutral graphene in the quantum regime, where one has $n_F$=0 (or -1) independent of the magnetic field, the cyclotron energy exhibits characteristic dependence $\Delta_{0(-1)}$$\propto$$\sqrt{B}$ \cite{Sadowski,Jiang,Deacon}. The spectral weight of the UHM is primarily dominated by the CR, therefore it can be readily measured experimentally. For the 100nm-wide GNR array, the UHM is observed in an energy range of 300-610 cm$^{-1}$, which corresponds to $\lambda_{IR}$$\approx$16-33 $\mu$m and $\lambda_{IR} / \lambda_{pl}$$\approx$$\lambda_{IR} / 2W$$\approx$80-165. This is consistent with the theoretical prediction that strong reduction in plasmon wavelength (and thus high degree of optical field confinement) can be achieved in graphene, up to $\lambda_{IR} / \lambda_{pl}$$\sim$200 \cite{Jablan}.

The effect of plasmon energy interchange between adjacent ribbons, plasmon cross-talk, is predicted in dense GNR arrays with $W/d$$>$2 \cite{Christensen,Nikitin_12}. Magneto-transmission spectra measured on the 100nm-wide GNR arrays with $W/d$=1:1 and $W/d$=2:1 do not reveal much variation in the position of the absorption line (Fig. 1(c)), indicating that the plasmon cross-talk effect in our samples can be largely neglected.

In Fig. 1(b), we plot the normalized magneto-transmission spectra, $T(B)/T$($B$=0), taken on an un-patterned 2D graphene reference sample, 100nm- and 50nm-wide GNR arrays. The 2D reference sample exhibits the typical CR of MEG with the absorption line corresponding to the $L_{0(-1)}$$\to$$L_{1(0)}$ LL transition \cite{Sadowski,newnote}. This transition is also observed in all GNR array samples, and for the 200nm-wide GNR array it is visible at very low magnetic fields down to 0.2 T \cite{note2}. This observation allows us to estimate the Fermi energy and the mobility in our samples. According to Pauli's exclusion principle, the $L_{0(-1)}$$\to$$L_{1(0)}$ transition is blocked when the $n$=1 (-1) LL becomes fully occupied (depleted). This corresponds to $E_F$$\leq$17 meV. Here, we want to emphasize that the Fermi energy in our samples is more than one order of magnitude smaller than $E_F$ reported in previous works on graphene magnetoplasmons \cite{Crassee,Yan2}, where the CR is found in the classical regime and the cyclotron energy $\hbar\omega_c$$\propto$$B$. To estimate the mobility, we use the semiclassical condition $\mu B$$>$1, and obtain $\mu$$>$50,000 cm$^2$V$^{-1}$s$^{-1}$ \cite{Orlita}. Such a high value of mobility suggests that MEG is an ideal system for studying graphene magnetoplasmons.

As one can also see in Fig. 1(b), the absorption line of GNR array samples is clearly blue-shifted with respect to the CR of 2D graphene, due to the formation of UHM. The energy shift increases as the GNR width decreases, and it becomes more pronounced at a lower magnetic field. This behavior is expected from Eq. (1), given that $\omega_{pl}$$\propto$$W^{-1/2}$. In addition, the amplitude of the absorption lines decreases with the GNR width (not shown in Fig. 1(b) where scaled spectra are plotted for clarity), even though the GNR width/gap ratio is the same,  $W/d$=1:1. This reduction in spectral weight prevents us from observing the UHM in narrow GNRs at very low mangetic fields.

\begin{figure}[t!]
\includegraphics[width=7cm] {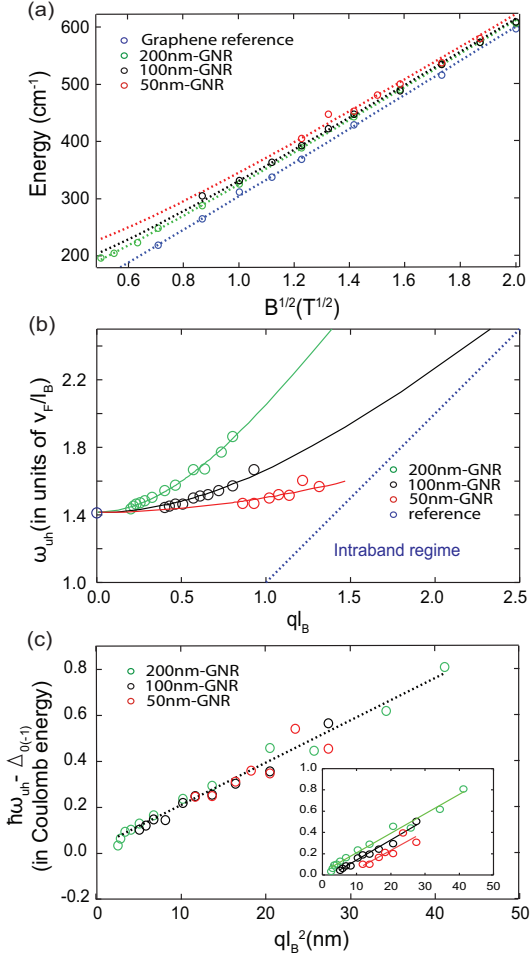}
\caption{(color online) (a) $\hbar\omega_{uh}$ versus $\sqrt{B}$ for 2D graphene, 200nm-, 100nm-, and 50nm-wide GNR array samples. Dotted lines represent best fits using Eq. (1). (b) Dispersion relation of the UHM, $\omega_{uh}(ql_B)$, in units of $v_F/l_B$. The data taken on the 2D graphene collapse on a single point (blue circle) at $q$=0. Solid lines show best fits to the data using Eq. (1) with $\omega_{pl}$ as fitting parameter, and dotted blue line indicates the boundary between the interband and intraband regimes. (c) Inset: The observed energy shift in GNR array samples with respect to the CR of 2D graphene, in units of Coulomb energy $e^2/(4\pi \epsilon \epsilon_0 l_B)$, as a function of $ql_B^2$. Main panel: Apart from a small offset, the data of the inset collapse on a single line. Dotted and solid lines represent best linear fits.}
\end{figure}

To examine the magnetic-field dependence of the UHM energies, we fit the transmission minima as a single Lorentzian and plot the extracted line positions as a function of $\sqrt{B}$ in Fig. 2(a). For all studied GNR array samples ($W$=200, 100, and 50 nm), the $\hbar\omega_{uh}(B)$ dependence is well described by Eq. (1) with $\Delta_{n_F}$=$\Delta_{0(-1)}$=$\sqrt{2}\hbar v_F/l_B$ for the $L_{0(-1)}$$\to$$L_{1(0)}$ LL transition. Here, $v_F$=(1.02$\pm$0.01)$\times$10$^6$ m/s is the Fermi velocity measured on the 2D reference sample \cite{Sadowski}, $l_B$=$\sqrt{\hbar/eB}$ is the magnetic length, and $e$ is the electron charge. The extracted values of $\hbar\omega_{pl}$ are summarized in Table I. Similar values of plasmon energies measured on two 100nm-wide GNR array samples with different thicknesses indicate that plasmon properties in MEG grown on the C-face of SiC can be considered similar to those in a stack of uncoupled graphene monolayers.

Since the UHM frequency is dominated by the contribution from CR, it is instructive to plot $\omega_{uh}$, in units of $v_F/l_B$, as a function of $ql_B$ (Fig. 2(b)). In this way, the CR line in Fig. 2(a) collapses on a single point at $q$=0 and the dispersion of the UHM is highlighted. The solid lines in Fig. 2(b) show best fits to the data using Eq. (1), from which we expect $\hbar\omega_{uh}$$\approx$$\Delta_{0}$ in the high magnetic field limit ($l_B$$\to$0). In the low field limit, Eq. (1) becomes insufficient to describe graphene magnetoplasmons as the UHM approaches the diagonal line $\omega$=$v_Fq$, separating regions of interband and intraband excitations. Contrary to conventional 2DEGs, the graphene plasmon frequency is $\omega_{pl}$$>$$v_F q$, for any value of $q$. The UHM can therefore never enter the intraband particle-hole continuum described by frequencies $\omega$$<$$v_F q$, but only the interband continuum, where it is weakly Landau-damped and merges into one of the linear magnetoplasmons \cite{Goerbig_prb,Hwang}.

\begin{table*}
\caption{Plasmon energies in GNR arrays extracted from the data of Figs. 2 and 3.}
\begin{ruledtabular}
\begin{tabular}{lrccc}
$\hbar\omega_{pl}$(cm$^{-1}$) & width/thickness: 200nm/72\AA & 100nm/128\AA	&	100nm/104\AA & 50nm/61\AA \\
\hline
un-polarized & 126\ \ \ \ \ \ \  & 149 & 143 & 172 \\
perpendicular $E$$||$$x$ & 154\ \ \ \ \ \ \  & 171 & X & X \\
parallel $E$$||$$y$ & 104\ \ \ \ \ \ \  & 133	& X & X \\
\end{tabular}
\end{ruledtabular}
\end{table*}

The UHM dispersion relation can be investigated further by considering the UHM energy shift with respect to the CR of 2D graphene. In the limit $\hbar\omega_{pl}$$\ll$$\Delta_{0}$, one can Taylor-expand Eq. (1) and express the energy shift in units of Coulomb energy as
\begin{equation}
\frac{\hbar\omega_{uh}-\Delta_{n_F=0}}{e^2/4\pi \epsilon_0\epsilon l_B}\simeq \frac{E_F}{\sqrt{2}\hbar v_F}q l_B^2\propto \frac{1}{WB},
\end{equation}
where $\epsilon_0$ is the vacuum permittivity, $\epsilon$=$(\epsilon_{SiC}+1)/2$$\approx$5 is the relative permittivity of epitaxial graphene, and the energy of the plasmon mode is given by
\begin{equation}
\hbar\omega_{pl}(q)=\sqrt{\frac{2e^2 E_F}{4\pi \epsilon_0\epsilon}q}.
\end{equation}
The above-mentioned Taylor expansion is valid if the plasmon term is small compared to the cyclotron energy. In our case,
\begin{equation}
\frac{\omega_{pl}^2(q)}{2v_F^2/l_B^2}=\frac{\alpha_G E_F}{\hbar v_F}q l_B^2\lesssim \alpha_G< 1,
\end{equation}
where $\alpha_G$=$2.2/\epsilon$ is the graphene fine structure constant. Therefore, Eq. (2) holds within our experimental parameter range and measurement accuracy.

\begin{figure}[b]
\includegraphics[width=8cm]{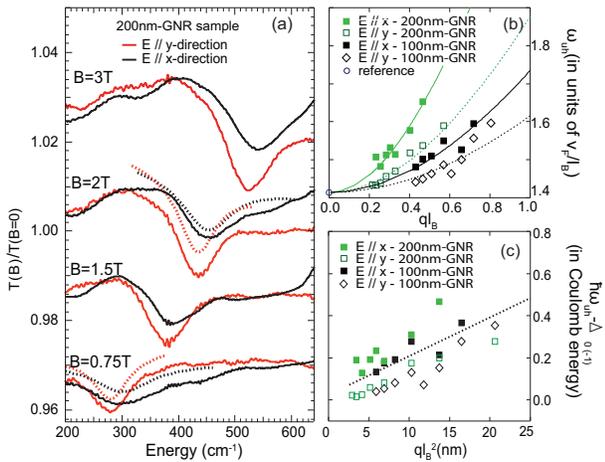}
\caption{(color online) (a) Normalized magneto-transmission spectra, $T(B)/T$($B$=0), of 200nm-wide GNR array with the IR light polarization parallel (red) or perpendicular (black) to the ribbon direction. Dotted lines are examples of Lorentzian fits to the data. The spectra and the fits are offset vertically for clarity. (b) Dispersion relation of the UHM, $\omega_{uh}(ql_B)$, in units of $v_F/l_B$ for both polarizations. Solid and dotted lines represent best fits using Eq. (1). (c) The UHM energy shift, in units of Coulomb energy, as a function of $ql_B^2$. Dotted guideline shows the $ql_B^2$ scaling with $E_F$=17 meV.}
\end{figure}

Note that the scaling behavior ($\propto$$ql_B^2$) in Eq. (2) is unique to the UHM in graphene in the quantum regime, while $\propto$$ql_B^3$ scaling is expected for conventional 2DEGs and highly doped graphene, where classical CR is observed. Moreover, this scaling behavior helps to distinguish the UHM from other possible interpretations of our data such as magnetoexcitons \cite{Iyengar}. Magnetoexcitons may be understood as inter-LL transitions acquiring a weak dispersion, as a function of the electron-hole wave vector, due to the mutual Coulomb interaction. For small values of $ql_B$$\ll$1 the magnetoexciton-induced energy shift scales as $\propto$$ql_B$. As shown in Fig. 2(c), the energy shift of the UHM in all studied GNR array samples scales as $ql_B^2$. Apart from a small vertical offset ($\sim$1.5\% of the cyclotron energy at $B$=4 T for the 100nm-wide GNR array), all data collapse on a single line with the slope corresponding to $E_F$=17$\pm2$ meV, in very good agreement with the estimated Fermi energy. One shall note that the presence of a vertical offset does not follow from Eq. (2). In our analysis, we considered a simple model using  the dispersion relation of plasmon mode for 2D graphene (Eq. (3)) and replacing the wave vector by $q$=$\pi/W$ to account for the GNR geometry.  Although this model provides correct insight about the dispersion and predicts $ql_B^2$ scaling, further corrective geometric terms, as those discussed in Ref. \cite{Christensen}, may give rise to the small offset in the data taken from GNR arrays with different widths.

Finally, we discuss the polarization-resolved measurements. Figure 3(a) shows a set of spectra taken with IR light polarized parallel ($E$$||$$y$) or perpendicular ($E$$||$$x$) to the ribbon direction. The absorption minima have essentially a Lorentzian lineshape with a slightly tilted baseline. For both 200nm- and 100nm-wide GNR arrays, the energy of the UHM absorption lines in $E$$||$$x$ polarization is blue-shifted and their linewidth is much broader compared with those for $E$$||$$y$ polarization. Figure 3(b) illustrates the obtained dispersion relation of the UHM for both polarizations and the best fits to the data using Eq. (1). The extracted values of $\hbar\omega_{pl}$ are summarized in Table I. Because of the IR intensity reduction caused by a linear polarizer and the weak CR absorption, polarization-resolved measurements were not possible for 50nm-wide GNR array. In Fig. 3(c), we examine the $ql_B^2$ scaling of the UHM energy shift with respect to the CR of 2D graphene. Interestingly, we find that $ql_B^2$ scaling with different vertical offsets holds for both polarizations. The difference in vertical offsets in polarized data is very similar to that found in un-polarized data for different GNR widths discussed previously, thus suggesting that additional geometric corrections to Eq. (2) depend on ribbons orientation with respect to the polarization direction. We notice that this polarization-dependent behavior of UHM was not observed in previously studied quasi-1D quantum wires \cite{Demel,Kern}. We hope our observation of surprising polarization effects in GNRs will trigger further theoretical studies which are beyond the scope of this experimental work.

In conclusion, we fabricated large-scale epitaxial GNR arrays with various widths and studied their plasmon properties in a magnetic field. We show that large-$q$ graphene plasmons can couple with the CR, forming an UHM and resulting in a blue-shift in the energy of inter-LL transition. The observed energy shift exhibits a peculiar $ql_B^2$ scaling, which distinguishes it from the UHM in conventional 2DEGs and in highly doped graphene.

The IR measurement of this work was supported by the DOE (DE-FG02-07ER46451), and the GNR array fabrication was supported by the NSF (DMR-0820382). The work at SNL was supported by the DOE Office of Basic Energy Sciences, Division of Materials Science and Engineering, and by Sandia LDRD. The work at GaTech was partially supported by the DOE BES through a contract with SNL. NHMFL is supported by the NSF (DMR-0654118), by the State of Florida, and by the DOE. Sandia National Laboratories is a multi-program laboratory managed and operated by Sandia Corporation, a wholly owned subsidiary of Lockheed Martin Corporation, for the United States Department of Energy's National Nuclear Security Administration under contract DE-AC04-94AL85000.


\begin{thebibliography}{99}
\bibitem{Ju}L. Ju \etal, Nature Nanotechnol. \textbf{6}, 630 (2011).
\bibitem{Xia}H. G. Yan \etal, Nature Nanotechnol. \textbf{7}, 330 (2012).
\bibitem{Crassee}I. Crassee \etal, Nano Lett. \textbf{12}, 2470 (2012).
\bibitem{Koppens_N12}J. N. Chen \etal, Nature \textbf{487}, 77 (2012).
\bibitem{Basov_N12}Z. Fei \etal, Nature \textbf{487}, 82 (2012).
\bibitem{Novoselov}A. N. Grigorenko, M. Polini, and K. S. Novoselov, Nature Photonics \textbf{6}, 749 (2012).
\bibitem{Jablan}M. Jablan, H. Buljan, and M. Solja\v{c}i\'{c}, Phys. Rev. B \textbf{80}, 245435 (2009).
\bibitem{Koppens}F. H. L. Koppens, D. E. Chang, and F. J. G. de Abajo, Nano Lett. \textbf{11}, 3370 (2011).
\bibitem{Engheta}A. Vakil and N. Engheta, Science \textbf{332}, 1291 (2011).
\bibitem{deHeer}M. Sprinkle \etal, Nature Nanotechnol. \textbf{5}, 727 (2010).
\bibitem{Pan}W. Pan et al., Appl. Phys. Lett. \textbf{97}, 252101 (2010).
\bibitem{Park}M. P. Levendorf \etal Nature \textbf{488}, 627 (2012).
\bibitem{Yan1}H. G. Yan \etal, arXiv:1209.1984.
\bibitem{Yan2}H. G. Yan \etal, Nano Lett. \textbf{12}, 3766 (2012).
\bibitem{Neto12}For a recent review see, for example, V. N. Kotov \etal, Rev. Mod. Phys. \textbf{84}, 1067 (2012).
\bibitem{Sadowski}M. L. Sadowski \etal, Phys. Rev. Lett. \textbf{97}, 266405 (2006).
\bibitem{Hass}J. Hass, F \etal, Phys. Rev. Lett. \textbf{100}, 125504 (2008).
\bibitem{Orlita}M. Orlita \etal, Phys. Rev. Lett. \textbf{101}, 267601 (2008).
\bibitem{Sprinkle}M. Sprinkle \etal, Phys. Rev. Lett. \textbf{103}, 226803 (2009).
\bibitem{note_Raman}Raman spectroscopy measurements confirm that 1-2 layers of graphene were formed on the Si-face during the epitaxial graphitization process.
\bibitem{Chiu}K. W. Chiu and J. J. Quinn, Phys. Rev. B \textbf{9}, 4724 (1974).
\bibitem{Demel}T. Demel \etal, Phys. Rev. B \textbf{38}, 12732 (1988).
\bibitem{Goerbig_11}R. Rold\'{a}n, M. O. Goerbig, and J.-N. Fuchs, Phys. Rev. B \textbf{83}, 205406 (2011); M. O. Goerbig, Rev. Mod. Phys. {\bf 83}, 1193 (2011).
\bibitem{Jiang}Z. Jiang \etal, Phys. Rev. Lett. \textbf{98}, 197403 (2007).
\bibitem{Deacon}R. S. Deacon \etal, Phys. Rev. B \textbf{76}, 081406(R) (2007).
\bibitem{Christensen}J. Christensen \etal, ACS Nano \textbf{6}, 431 (2012).
\bibitem{Nikitin_12}A. Yu. Nikitin \etal, Phys. Rev. B \textbf{85}, 081405 (2012).
\bibitem{newnote}The graphene grown on the back side (Si-face) is highly doped and has much lower mobility, compared with the quasi-neutral MEG on the C-face. Therefore, it will not affect the LL transition spectra in the energy window of this work for the magnetic field applied \cite{Witowski}.
\bibitem{Witowski}A. M. Witowski \etal, Phys. Rev. B \textbf{82}, 165305(2010).
\bibitem{note2}For the 100nm- and 50nm-wide GNR arrays, the $L_{0(-1)}$$\to$$L_{1(0)}$ LL transition vanishes when the magnetic length becomes comparable with the ribbon width.
\bibitem{Goerbig_prb}R. Rold\'{a}n, J.-N. Fuchs, and M. O. Goerbig, Phys. Rev. B \textbf{80}, 085408 (2009);
R. Rold\'{a}n, M. O. Goerbig, and J.-N. Fuchs, Semicond. Sci. Technol. \textbf{25}, 034005 (2010).
\bibitem{Hwang}E. H. Hwang and S. Das Sarma, Phys. Rev. B \textbf{80}, 205405 (2009).
\bibitem{Iyengar}A. Iyengar, J. Wang, H. A. Fertig, and L. Brey, Phys. Rev. B \textbf{75}, 125430 (2007);
K. Shizuya, Phys. Rev. B \textbf{75}, 245417 (2007); Y. A. Bychkov and G. Martinez, Phys. Rev. B \textbf{77}, 125417 (2008);
R. Rold\'{a}n, J.-N. Fuchs, and M. O. Goerbig, Phys. Rev. B \textbf{82}, 205418 (2010).
\bibitem{Kern}K. Kern \etal, Surface Science \textbf{229}, 256 (1990).
\end{thebibliography}
\end{document}